\documentclass[11pt,a4paper]{article}
\pdfoutput=1

\usepackage{epsf}

\usepackage{epsfig,graphics}
\usepackage{graphicx}
\usepackage{epsfig}
\usepackage{subfigure}
\usepackage{tabularx} 
\usepackage{rotate}	
\usepackage{slashed}
\usepackage{color}

\usepackage{amsmath}
\usepackage{amsfonts}
\usepackage{amssymb}
\usepackage{graphicx}
\usepackage{cite}

\newcommand{\bmat}{\left(\begin{array}}
\newcommand{\emat}{\end{array}\right)}

\def\yzero{\smash{\hbox{$y\kern-4pt\raise1pt\hbox{${}^\circ$}$}}}

\def\beq{\begin{equation}}
\def\eeq{\end{equation}}
\def\beqa{\begin{eqnarray}}
\def\eeqa{\end{eqnarray}}

\def\-{\hphantom{-}}

\def\s2{\frac{1}{\sqrt2}}

\def\oh{\frac{1}{2}}
\def\beq{\begin{equation}}
\def\eeq{\end{equation}}
\def\beqa{\begin{eqnarray}}
\def\eeqa{\end{eqnarray}}

\def\IF{\relax{\rm I\kern-.18em F}}
\def\II{\relax{\rm I\kern-.18em I}}
\def\IP{\relax{\rm I\kern-.18em P}}
\def\IC{\relax\hbox{\kern.25em$\inbar\kern-.3em{\rm C}$}}
\def\IR{\relax{\rm I\kern-.18em R}}

\def\Dsl{\,\raise.15ex\hbox{/}\mkern-13.5mu D} 
\def\IZ{Z\kern-.4em  Z}

\def\-{\hphantom{-}}

\def\s2{\frac{1}{\sqrt2}}

\def\oh{\frac{1}{2}}

\def\IC{\mathbb{C}}
\def\sig{{\sigma}}

\def\CF {{\cal F}}

\def\IF{\relax{\rm I\kern-.18em F}}
\def\II{\relax{\rm I\kern-.18em I}}
\def\IP{\relax{\rm I\kern-.18em P}}
\def\IC{\relax\hbox{\kern.25em$\inbar\kern-.3em{\rm C}$}}
\def\IR{\relax{\rm I\kern-.18em R}}

\newcommand{\be}{\begin{equation}} \newcommand{\ee}{\end{equation}}
\newcommand{\bea}{\begin{eqnarray}} \newcommand{\eea}{\end{eqnarray}}
\newcommand{\beann}{\begin{eqnarray*}}  \newcommand{\eeann}{\end{eqnarray*}}
\newcommand{\bfig}{\begin{figure}} \newcommand{\efig}{\end{figure}}
\newcommand{\ba}{\begin{array}} \newcommand{\ea}{\end{array}}
\newcommand{\bcen}{\begin{center}} \newcommand{\ecen}{\end{center}}
\newcommand{\btab}{\begin{tabular}} \newcommand{\etab}{\end{tabular}}

\newcommand{\matt}{\left ( \begin{array}{ccc}}
\newcommand{\ematt}{\end{array} \right )} \newcommand{\matf}{\left
(\begin{array}{cccc}}

\catcode`\@=11   
\newdimen\@rotdimen
\newbox\@rotbox  

\def\@vspec#1{\special{ps:#1}}
\def\@rotstart#1{\@vspec{gsave currentpoint currentpoint translate
   #1 neg exch neg exch translate}}
\def\@rotfinish{\@vspec{currentpoint grestore moveto}}
\def\@rotr#1{\@rotdimen=\ht#1\advance\@rotdimen by\dp#1%
   \hbox to\@rotdimen{\hskip\ht#1\vbox to\wd#1{\@rotstart{90 rotate}%
   \box#1\vss}\hss}\@rotfinish}
\def\@rotl#1{\@rotdimen=\ht#1\advance\@rotdimen by\dp#1%
   \hbox to\@rotdimen{\vbox to\wd#1{\vskip\wd#1\@rotstart{270 rotate}%
   \box#1\vss}\hss}\@rotfinish}%
\def\@rotu#1{\@rotdimen=\ht#1\advance\@rotdimen by\dp#1%
   \hbox to\wd#1{\hskip\wd#1\vbox to\@rotdimen{\vskip\@rotdimen
   \@rotstart{-1 dup scale}\box#1\vss}\hss}\@rotfinish}%
\def\@rotf#1{\hbox to\wd#1{\hskip\wd#1\@rotstart{-1 1 scale}%
   \box#1\hss}\@rotfinish}%
\def\rotate{\@ifnextchar[{\@rotate}{\@rotate[l]}}
\def\@rotate[#1]#2{\setbox\@rotbox=\hbox{#2}\@nameuse{@rot#1}\@rotbox}

\catcode`\@=12
\begin{document}

\makeatletter
\@addtoreset{equation}{section}
\makeatother
\renewcommand{\theequation}{\thesection.\arabic{equation}}


\pagestyle{empty}
\rightline{ IFT-UAM/CSIC-16-007}
\rightline{ MPP-2016-6}
\vspace{1.2cm}


\vskip 1.5cm

\begin{center}
\LARGE{The DBI Action, Higher-derivative Supergravity, \\ and Flattening Inflaton Potentials} 
\\[13mm]
  \large{Sjoerd Bielleman$^{1}$,  Luis E. Ib\'a\~nez$^{1}$, Francisco G. Pedro$^{1}$, \\ Irene Valenzuela$^{2,3}$, and Clemens Wieck$^{1}$  \\[6mm]}
\small{
${}^1$  Departamento de F\'{\i}sica Te\'orica
and Instituto de F\'{\i}sica Te\'orica UAM/CSIC,\\[-0.3em]
Universidad Aut\'onoma de Madrid,
Cantoblanco, 28049 Madrid, Spain\\
${}^2$ Max-Planck-Institut fur Physik,
Fohringer Ring 6, 80805 Munich, Germany\\
${}^3$Institute for Theoretical Physics and
Center for Extreme Matter and Emergent Phenomena,\\
Utrecht University, Leuvenlaan 4, 3584 CE Utrecht, The Netherlands
\\[8mm]}
\end{center}
\begin{abstract}
In string theory compactifications it is common to find an effective Lagrangian for the scalar fields with a non-canonical kinetic term. We study the effective action of the scalar position moduli of Type II D$p$-branes. In many instances the kinetic terms are in fact modified by a term proportional to the scalar potential itself. This can be linked to the appearance of higher-dimensional supersymmetric operators correcting the K\"ahler potential. We identify the supersymmetric dimension-eight operators describing the $\alpha'$ corrections captured by the D-brane Dirac-Born-Infeld action. Our analysis then allows an embedding of the D-brane moduli effective action into an $\mathcal N = 1$ supergravity formulation. The effects of the potential-dependent kinetic terms may be very important if one of the scalars is the inflaton, since they lead to a flattening of the scalar potential. We analyze this flattening effect in detail and compute its impact on the CMB observables for single-field inflation with monomial potentials.\end{abstract}
\newpage
\setcounter{page}{1}
\pagestyle{plain}
\renewcommand{\thefootnote}{\arabic{footnote}}
\setcounter{footnote}{0}



\tableofcontents

%

\section{Introduction}

Renormalizable field theories like the Standard Model only include up to two derivatives in the action. However, gravitational interactions and unified schemes like string theory which go beyond the Standard Model do contain higher-order derivative couplings. More generally, they contain higher-dimensional operators suppressed by the cut-off scale of the theory. Higher-dimensional operators involving scalars are potentially important when studying the vacua and the cosmological evolution of a theory. For example, in cosmic inflation the inflaton field may start slow-roll at very large classical values and the kinetic terms may be non-minimal. Another situation in which such operators may be important is in moduli stabilization in string vacua with vacuum expectation values close to the string or Planck scale.

Higher-dimensional operators involving chiral superfields have been studied in the past in supersymmetry and supergravity \cite{Cecotti:1986jy,Cecotti:1986pk,Antoniadis:2007xc,Khoury:2010gb,Baumann:2011nm,Koehn:2012ar,Farakos:2012qu,Koehn:2012np,Koehn:2012te,Farakos:2013zsa,Gwyn:2014wna,Aoki:2015eba,Ciupke:2015msa}. In general the K\"ahler potential may depend on the superfields and their derivatives, i.e.,
\begin{align}
\int d \theta^2 d {\bar \theta }^2 K(\Phi_i, \bar \Phi_i;  D_\alpha \Phi_i, \bar D_{\dot \alpha} \bar \Phi_i ; \partial_\mu \Phi_i, \partial_\mu \bar \Phi_i, \dots)  + \left( \int d\theta^2 W(\Phi_i) +\text{h.c.} \right)\,,
\end{align}
where $D_\alpha$ denotes the usual supersymmetric covariant derivative. Higher-dimensional corrections to $W$ are generally model-dependent and involve, for example, higher powers of the superfields. Therefore we focus on generic corrections to the D-term. One usually expands the action in fields and covariant derivatives, keeping only the leading contribution of the higher-dimensional operators. Such operators may lead to problems if not appropriately constrained. In particular, upon expansion in components ghosts may appear and/or the auxiliary fields of the superfields may become propagating.

A classification of such operators was presented in \cite{Khoury:2010gb}. A particular ghost-free linear combination of them has been singled out, it reads
\beq
\int d\theta^2  d{\bar \theta}^2 D^\alpha \Phi D_\alpha \Phi \bar D_{\dot \alpha} \bar \Phi \bar D^{\dot \alpha} \bar \Phi  \,.
\label{phi4}
\eeq
This operator and its component expression are simple and ``clean'' for a number of reasons, and it has the potential advantage that it can be coupled to $\mathcal N = 1$ supergravity in a straight-forward manner \cite{Koehn:2012te}. Unfortunately, as discussed below, it is not the operator we find in the effective action of D-branes in Type II string theory. Notice that, since $D^\alpha \Phi= 2\theta^\alpha F + \dots$ where $F$ is the auxiliary field, \eqref{phi4} includes a term proportional to $|F|^4$. This means that the equation of motion for $F$ is cubic and has three solutions. This fact has been discussed in \cite{Khoury:2010gb} and more recently in \cite{Ciupke:2015msa} and \cite{Broy:2015zba}, where the operator was applied to K\"ahler moduli stabilization and inflation in Type IIB string compactifications. 

In this paper we consider the issue of higher-derivative operators from the point of view of the effective action of string theory. In particular, we study the effective action for the scalars corresponding to position moduli of Type II D$p$-branes. Such scalars parameterize the motion of D$p$-branes in compact dimensions and have been considered as possible inflaton candidates in many models of string inflation \cite{Hebecker:2014eua,Ibanez:2014kia,Arends:2014qca,Ibanez:2014swa,Bielleman:2015lka}, cf.~\cite{Baumann:2014nda} for an exhaustive list of references. The bosonic action is given by the non-Abelian generalization of the Dirac-Born-Infeld (DBI) action and the Chern-Simons (CS) action \cite{Ibanez:2012zz}. The former captures all higher-dimensional operators involving arbitrary powers of single derivatives and the scalars themselves. Hence it can give us information about higher-dimensional supersymmetric operators in the effective action. We analyze the effective actions of the D-brane moduli of Type IIB D$p$-branes for $p=3,5,7$. As mentioned above, we find that the operator in (\ref{phi4}) in fact never appears in these actions. Instead we find operators of the form
\beq
\int  d\theta^2  d{\bar \theta}^2 |\Phi|^2 \partial_\mu \Phi \partial^\mu \bar \Phi \,,
\eeq
and variations thereof. An important property of this class of operators is that no terms proportional to $|F|^4$ arise and hence the solution of the equations of motion for $F$ is unambiguous. On the other hand, there appear non-canonical kinetic terms proportional to $(1+|F|^2)(\partial_\mu \phi \partial^\mu \bar \phi)$, where $\phi$ denotes the complex scalar component of $\Phi$. This matches what we find in the string-effective action: in orientifold compactifications, the kinetic Lagrangian of the D-brane position moduli $\phi_i$ has the on-shell structure
\begin{align}
\mathcal{L} =- \left[1+a V(\phi_i)\right] \partial_\mu \phi_i \partial^\mu \bar \phi_i - V(\phi_i) \,,
\label{higgsotic}
\end{align}
where $V$ is in many cases the leading-order scalar potential and $a$ is a constant proportional to the inverse fourth power of the string scale $M_s=(\alpha')^{-1/2}$. This result is exact at second order in the derivatives and at all orders in the potential. In particular, no corrections of the form $V^n$ with $n>1$ arise. Describing the string-effective DBI action in terms of supersymmetric higher-derivative operators allows an embedding into an $\mathcal N=1$ supergravity formulation. In \cite{Ibanez:2014swa}, for example, a supergravity description was proposed which did not account for the higher-derivative terms. We can now close this gap by including the above operators, which allows us to study the flattening effects in a supergravity formulation of \cite{Ibanez:2014swa} in combination with, for example, closed string moduli stabilization.

It is clear that the higher-dimensional terms induce a non-canonical redefinition of the kinetic terms, which leads to a flattening of the effective scalar potential. A similar redefinition of kinetic terms was discussed in string inflation in \cite{Gur-Ari:2013sba}. In a second part of this paper we analyze the consequences for the inflationary dynamics. We give general analytic formulae for the slow-roll parameters modified by the non-canonical kinetic terms in \eqref{higgsotic}, focussing on monomial inflaton potentials. In all cases the non-canonical kinetic term leads to a flattening of the potential at large field values. This causes a substantial reduction of the tensor-to-scalar ratio $r$, bringing chaotic inflation models to better agreement with the recent Planck and BICEP data.

The structure of this paper is as follows. In the next section we study the structure of the effective action for D$p$-brane moduli in
Type IIB toroidal compactifications. We analyze in detail the cases of D3-, D5-, and D7-branes and display the bosonic action up to fourth order in derivatives. The result is always of the form \eqref{higgsotic}. In Section 3 we discuss higher-derivative operators in globally supersymmetric theories in general and describe how the result obtained from the DBI action can be written in terms of these operators. Moreover, we show how these operators lead to a supergravity description of the flattening effect in D-brane models like the one of \cite{Ibanez:2014swa}, and comment on closed string moduli stabilization. In Section 4 we use the structure in \eqref{higgsotic} applied to a single inflaton field to study the behavior of the slow-roll parameters for varying values of $a$. Section 5 is left for our conclusions. 

%

\section{Higher-derivative terms for D-brane moduli from the DBI action}\label{sec:DBI}

The four-dimensional effective theory for the bosonic open string fields of D$p$-branes can be derived from the DBI and CS actions describing the world-volume deformations of the brane.  This is especially useful in the case of toroidal compactifications, in which the internal profile of the scalar fields is constant and the compactification to four dimensions is trivial. The DBI action is exact in $\alpha'$ up to second derivatives of the scalars, leading to a clear advantage over the standard supergravity description of the effective theory for open string moduli in which $\alpha'$ corrections are in general not known or highly difficult to compute.\footnote{Cf.~\cite{Grimm:2014xva,Grimm:2014efa,Grimm:2015mua} for recent studies of $\alpha'$ corrections and higher-derivative terms on M-theory reductions.} We find that these $\alpha'$ corrections affect the kinetic term, giving rise to a non-canonical normalization as advanced in the Introduction. Keeping track of these corrections, though interesting by itself, is essential in the study of large-field inflation models. In this section we study under which circumstances the schematic structure  \eqref{higgsotic} arises for the open string fields of a system of D$p$-branes in Type IIB orientifold compactifications, leaving the inflationary analysis to Section \ref{sec:inflation}.

Let us start by giving the general form of the DBI action for D$p$-branes \cite{Tseytlin:1997csa,Myers:1999ps,Ibanez:2012zz},
\beq
S=  -\mu_p \int d^{p+1} \xi \, {\rm STr} \, e^{-\phi}
\sqrt{-\det(P[E_{MN}+E_{M i}(Q^{-1}-\delta)^{ij}E_{jN}]
+\, \sigma F_{MN})\, \det(Q_{mn})}  \, .
\eeq
The integral goes over the $(p+1)$-dimensional world-volume of the D$p$-brane and we have defined as usual
\begin{align}
\sigma = 2\pi \alpha' \,, \qquad E_{MN} = g_s^{1/2}G_{MN}- B_{MN} \,, \qquad Q_{mn}  = \delta_{mn}+i\sigma [\varphi_m, \varphi_p]E_{pn}\,,
\end{align}
$M$ and $N$ are ten-dimensional indices, $\mu$ and $\nu$ are spacetime indices, $a$ and $b$ are internal indices labelling the $(p-3)$-cycle wrapped by the brane, and $m$ and $n$ label the real coordinates transverse to the brane. The $\varphi_m$ are the real position moduli. We consider an ansatz for the metric given by
\beq
 ds^2=Z(x^m)^{-1/2}\eta_{\mu\nu} dx^{\mu} dx^{\nu}+Z(x^m)^{1/2} ds^2_\text{CY}\,,
\eeq
where $Z$ denotes a possible warp factor and $ds^2_\text{CY}$ the line element in the internal Calabi-Yau manifold.

We focus our discussion on the position moduli of the D$p$-branes because they are the ones which get a potential once fluxes are added. Thus, we omit all terms involving gauge bosons and Wilson lines.\footnote{Cf.~\cite{Escobar:2015fda,Escobar:2015ckf} for a recent analysis of inflation with D6-brane Wilson lines in type IIA.} In the absence of mixed Minkowski-internal tensors, i.e., $g_{\mu a}=B_{\mu a}=0$, and considering a constant internal profile for the position moduli, $\partial_a\phi=0$, the world-volume determinant can be factorized as
\begin{align}
&\det(P[E_{MN}+ \sigma F_{MN}]) \nonumber \\ 
&\;\;\;\; = \det(g_s^{1/2}Z^{-1/2}\eta_{\mu\nu}+g_s^{1/2}Z^{1/2}\sigma^2\partial_\mu \varphi_m\partial_\nu \varphi_n)\det(g_s^{1/2}g_{ab} +\sigma { F}_{ab}-B_{ab})\,.
\end{align}
This factorization of Minkowski and internal indices is exact in toroidal compactifications. However, in a Calabi-Yau compactification the internal profile of the scalar fields is in general not constant. This implies one has to solve an eigenstate equation for the internal space, which is usually non-trivial. Besides, the zero eigenmodes might correspond to mixings between the original position moduli and Wilson lines, making the computation technically much more involved. Therefore we restrict our study to the simplest cases in which the above factorization can be performed. For a D3-brane all world-volume indices are in Minkowski spacetime so there are no subtleties regarding the compactification.

Moreover, taking into account the contribution from the transverse coordinates, the quantity inside the square root in the DBI action is composed of three factorized determinants,
\begin{subequations}
\label{factor}
\begin{eqnarray}
&&\det(g_s^{1/2}Z^{-1/2}\eta_{\mu\nu}+g_s^{1/2}Z^{1/2}\sigma^2\partial_\mu \varphi_m\partial_\nu \varphi_m)\,,\\ 
&&\det(g_s^{1/2}g_{ab} +\sigma { F}_{ab}-B_{ab})\,, \\ 
&&\det(g_{mn}+i\sigma [\varphi_m,\varphi_p](g_s^{1/2}g_{pn}-B_{pn}))\,.
\end{eqnarray}
\end{subequations}
For a D$p$-brane these three matrices have dimension 4, $(p-3)$, and $(9-p)$, respectively. After rearranging the real fields $\varphi_m$ in a complex basis denoted by $\phi_i$, the first determinant becomes
\begin{align}\nonumber
&-\det(g_s^{1/2}Z^{-1/2}\eta_{\mu\nu} + g_s^{1/2}Z^{1/2}\sigma^2\partial_\mu \varphi_m\partial_\nu \varphi_m) \nonumber \\ 
& \hskip 0.5cm = g_s^{2}Z^{-2} \bigg( 1+2Z\sigma^2\partial^{\mu}\phi_i\partial_{\mu}\bar \phi_i + Z^2\sigma^4 \big[ 2(\partial^{\mu}\phi_i\partial_{\mu}\bar \phi_i)^2 \nonumber \\
& \hskip 3cm - (\partial^{\mu}\phi_i\partial_{\mu}\bar \phi_j)(\partial^{\nu}\phi_j\partial_{\nu}\bar \phi_i)-(\partial^{\mu}\phi_i\partial_{\mu} \phi_j)(\partial^{\nu}\bar\phi_i\partial_{\nu}\bar \phi_j) \big] \bigg) \,.
\end{align}
We can now Taylor-expand the square root in powers of spacetime derivatives of $\phi$. This expansion is 
in accordance with the slow-roll approximation during inflation. This yields
\begin{multline}
\mathcal{L}  = -\mu_p  Z^{-1} V_{p-3} f(\phi) \left(1+Z\sigma^2\sum_i\partial_\mu \phi_i\partial^\mu \bar\phi_i -  \frac{1}{2} Z^2\sigma^4\left[\sum_{i\neq j}(\partial^{\mu}\phi_i\partial_{\mu}\bar \phi_j)(\partial^{\nu}\phi_j\partial_{\nu}\bar \phi_i) \right.\right.\\
\left.\left.+\sum_{i,j}(\partial_\mu \phi_i\partial^\mu \phi_j)(\partial_\nu \bar\phi_i\partial^\nu \bar\phi_j) \right] + \dots \right)\,,
\label{L}
 \end{multline}
with 
\beq
f(\phi) = \sqrt{ \det(g_s^{1/2}g_{ab} +\sigma { F}_{ab}-B_{ab})
\det(g_{mn}+i\sigma [\varphi_m,\varphi_p](g_s^{1/2}g_{pn}-B_{pn}))}\,.
\label{f}
\eeq
Here $\mu_p$ and $V_{p-3}$ denote the tension of the brane and the volume wrapped by the brane, respectively. Note that, after the square root expansion, no term of the form $(\partial_\mu \phi \partial^\mu\bar\phi)^2$ is present in the effective action in the case of a single complex position modulus, i.e., for a D7-brane. 

We observe that in all cases the bosonic action has the structure 
\begin{align}
\mathcal{L} = -\left[1+a V(\phi) \right] |\partial_\mu \phi|^2 - V(\phi) + \mathcal{O}(\partial_\mu^4) \,, 
\label{higgsotic2}
\end{align}
where we have implicitly redefined the scalar fields to absorb the global factors in \eqref{L}. We have also subtracted the orientifold tension which is required for an (approximate) Minkowski vacuum, cf.~\cite{Marchesano:2014mla,Ibanez:2014swa}, implying that ${V(\phi)=a^{-1}(f(\phi)-1)}$. The constant $a$ includes the remaining global factors and is proportional to $(\mu_pV_{p-3})^{-1}$, so it has mass dimension $-4$. Let us remark that the above result includes all $\alpha'$ corrections arising from higher-order terms containing powers $\phi^n$ in the DBI action. However, it is an expansion in derivatives of the scalar fields, so it can only be trusted as long as they remain small compared to the string scale. In particular, the DBI action does not include information about second- or higher-order derivatives of $\phi$, which will be important in the next section.

In addition to the DBI piece discussed above there is a contribution from the CS action. As discussed in more detail in \cite{Ibanez:2014swa}, in supersymmetric settings this contribution is equal to the DBI piece, leading to a factor of two in front of the scalar potential in the above expression -- but not in the correction to the kinetic term.

The structure of the scalar potential depends, through the specific form of $f(\phi)$, on the D$p$-brane under consideration and on the closed string background. In the following we summarize the results for D7-, D3-, and D5-branes in Type IIB orientifold compactifications.

\begin{itemize}

\item D7-branes

In the case of D7-branes there is only one complex scalar field $\phi$ in the adjoint representation of the gauge group of the system of D7-branes. 
One can obtain more realistic quantum numbers, for example the Standard Model gauge group and bifundamentals, if the branes are located at orbifold singularities, cf.~\cite{Ibanez:2014swa}.
The scalar $\phi$ parameterizes the position of the brane in the two-real dimensional transverse space. In the presence of three-form closed string fluxes $G_3$, the position of the branes can be stabilized due to the flux-induced $B$-field on the brane which yields a non-vanishing F-term scalar potential for $\phi$. This potential comes from the first determinant in \eqref{f} which reads
\be
\det \left(g_{ab}+Z^{-1/2}g_s^{-1/2}\CF_{ab}\right) = \det (g_{ab}) \left[1 + Z^{-1}g_s^{-1}\CF^2  +Z^{-2}g_s^{-2} \frac{1}{4} \left(\CF \wedge \CF\right)^2\right]\,,
\label{DBIint}
\ee
where $\CF_{ab} = \sig F_{ab} - B_{ab}$. Whenever $\CF$ is a selfdual or anti-selfdual two-form, $\CF \, =\, \pm *_4 \CF$,
we have
\be
\left(\CF \wedge \CF\right)^2 = \left(\CF \wedge *_4 \CF\right)^2 = \left(\CF^2 d{\rm vol}_{S_4} \right)^2 = (\CF^2)^4 \,,
\ee
and hence
\be
f(\phi)^2 = g_s^2Z^2\left(1 + \oh Z^{-1}g_s^{-1} \CF_{ab}\CF^{ab}\right)^2\,,
\ee
a perfect square. This is the case for a configuration with only imaginary selfdual closed string fluxes including $(0,3)$-form and $(2,1)$-form fluxes denoted by $G$ and $S$, respectively
\cite{Grana:2002tu,Camara:2003ku,Grana:2003ek,Lust:2004fi,Camara:2004jj,Aparicio:2008wh,Camara:2014tba}. In that case the $B$-field is a $(2,0)+(0,2)$-form. Far from being isolated or useless cases, these are indeed the fluxes which solve the ten-dimensional supergravity equations of motion in a Calabi-Yau compactification \cite{Giddings:2001yu}. The F-term scalar potential, after a field redefinition, reads \cite{Ibanez:2014swa}
\beq
V(\phi)=\frac{Z^{-2}g_s}{2} |G^*\phi-S\bar\phi|^2 \,.
\label{VD7}
\eeq
In addition to this flux potential there is a contribution from the superpotential that couples the modulus to two complex Wilson line scalars, which we have omitted in the DBI reduction for simplicity. These two fields together with $\phi$ complete the scalar components of the $\mathcal N=4$ structure that underlies the toroidal compactification before any twist or background decreases the number of supersymmetry generators. On the other hand, the second determinant in \eqref{f} leads to a D-term given by
\beq
\det(Q_{ij})=1+g_s\sigma^2 Z[\phi,\bar\phi]^2\,.
\eeq
For simplicity we consider D-flat configurations and neglect this contribution to the scalar potential from now on. The generalization to non-vanishing D-terms is trivial and does not change any of our conclusions. Finally, the CS contribution to $V$ can be checked to be equal to \eqref{VD7} when only $G$ and $S$ fluxes are turned on. Therefore the effective Lagrangian is of the form \eqref{higgsotic2} with $V$ given by \eqref{VD7} and $a=\frac12 (V_4\mu_7g_s)^{-1}$. 

\item D3-branes

In the case of D3 branes only the second determinant in \eqref{f} is present since all world-volume indices are spacetime indices. Notice then that the structure \eqref{higgsotic2} is more robust than for the case of D7- or D5-branes because the factorization \eqref{factor} always exists, regardless of the specific compactification. At leading order the square of \eqref{f} is given by
\begin{align}\nonumber
\det(\delta_{mn}+i\sigma g_s^{1/2}Z^{1/2} [\varphi_m,\varphi_n]) &= 1-2\sigma^2g_sZ\sum_{i<j} [\phi_i,\phi_j]^2 -\sigma^2 g_sZ\sum_{i,j} [\phi_i,\bar \phi_j]^2 +\dots \\
&= 1+\sum_i|F_i|^2 +\sum_i D_i^2+\dots\,,
\end{align}
where the dots include higher-order terms in $\sigma$. Notice that at leading order this corresponds to the sum of three F-terms and three D-terms. It is remarkable that in the absence of D-terms the above determinant can again be written as a perfect square,
\be
f(\phi)^2=\det(\delta_{mn}+i\sigma g_s^{1/2}Z^{1/2}[\varphi_m,\varphi_n]) = \left(1-\sigma^2g_sZ\sum_{i<j} [\phi_i,\phi_j]^2\right)^2 \,,
\ee
implying \eqref{higgsotic2} with $a=\mu_3^{-1}Z$ and $V=\sum_{i<j} g_s[\phi_i,\phi_j]^2$. This structure is partially broken if we introduce warping and fluxes. The situation is slightly more subtle since, as described in \cite{Grana:2002tu,Camara:2003ku,Grana:2003ek,Lust:2004fi,Camara:2004jj,Aparicio:2008wh,Camara:2014tba}, the local equations of motion force the internal metric and five-form background to be non-vanishing. One can then locally expand the warp factor around the position of the brane as
\beq
Z^{-1/2}=Z^{-1/2}_0+\frac12 \sigma^2 K_{mn}\varphi^m\varphi^n+\dots \,.
\eeq
This induces an additional contribution to the scalar potential coming from the warp factor $Z$ in \eqref{L} which does not appear multiplying the kinetic term. Therefore, in the presence of non-constant warping the correction to the kinetic term is given by only a part of the scalar potential.

\item D5-branes

The result for D5-branes is a combination of the two cases considered above. Both determinants in \eqref{f} contribute to the F-term scalar potential. The computation is simple in a purely supersymmetric configuration with no D-terms or fluxes. In that case, 
\be
f(\phi)^2=\det(\delta_{mn}+i\sigma g_s^{1/2}Z^{1/2} [\varphi_m,\varphi_n]) = \left(1-4\sigma^2g_sZ[\phi_1,\phi_2]^2\right)^2 \,,
\ee
where $\phi_1$ and $\phi_2$ are the two complex fields parameterizing the position of the D5-brane in the transverse space, which we have assumed to be a $T^4$ for simplicity. We thus once more obtain a Lagrangian of the form \eqref{higgsotic2} with $a=\mu_5^{-1}V_2^{-1}g_s^{-1/2}Z^{1/2}$ and $V=(\mu_5V_2\sigma^2)^{-1}Z^{-1/2}g_s^{1/2}[\phi_1,\phi_2]^2$.

\end{itemize}

In general, expression \eqref{f} is an infinite series in powers of $\sigma$. However, we have seen that in  certain configurations the determinant is a perfect square, simplifying the computation. This is the case for configurations which preserve a certain amount of supersymmetry at the string scale, i.e., when the D-terms vanish and only specific choices of fluxes are allowed so that supersymmetry can be spontaneously -- not explicitly -- broken at a lower scale. In that case, taking the square root of the determinant is trivial and the scalar potential is given by the leading-order scalar potential $V_0$. In other words, all higher-order terms in $\alpha'$ vanish, so the potential is simply $V=V_0$. However, these corrections do leave a trace in the effective theory because the kinetic terms for the scalar fields are non-canonical. The prefactor of the kinetic term is indeed given by $(1+a V_0)$, where $a$ is a constant depending on the brane tension and the string scale, showing the stringy nature of the correction.

Let us stress that the structure \eqref{higgsotic2} is quite general and valid beyond the supersymmetric configurations described here as examples. The advantage of these configurations is that one can replace $V$ by the well-known leading-order result $V_0$ to simplify the computation, while in general the scalar potential receives corrections as well. However, those corrections will also appear in the kinetic term, implying that the structure \eqref{higgsotic2} is preserved anyway. In the case of D5 and D7-branes, this structure relies on the assumption that the factorization \eqref{factor} can be done, which is characteristic of toroidal compactifications. It would be interesting to study to what extent it can be generalized to more general compactifications.

Finally, notice that the scalar potential $V(\phi)$ entering in the non-canonical kinetic term is only the contribution from the DBI action and not the full potential in general. However, in the supersymmetric configurations described above, the CS contribution equals the DBI potential, and the prefactor $f(\phi)$ is indeed a function of the full scalar potential for $\phi$, including the pieces generated
by background fluxes. This is the case we have in mind in Section~\ref{sec:inflation} when studying the implications of this structure for inflation.

%

\section{Supersymmetric higher-derivative operators and the DBI action}
\label{sec:higherd}

The lesson of the previous discussion is that the DBI action yields a very particular four-dimensional effective action for the D-brane position moduli. For simplicity, let us consider the case of a single complex modulus because the generalization to an arbitrary number of open string moduli is straightforward. The action can be written as
\begin{align}
\mathcal L = -\left[ 1 + a V(\phi) \right] \partial_\mu \phi \partial^\mu \bar \phi + |\partial_\mu \phi \partial^\mu \phi |^2 - V(\phi)\,,
\label{dbiresult}
\end{align}
at four-derivative order and after absorbing all global coefficients. This corresponds, for example, to the case of D7-branes in a toroidal background. The aim of this section is to identify the supersymmetric higher-derivative operators which lead to the DBI result in \eqref{dbiresult}. The above correction to the kinetic term is purely of stringy nature. Hence it can be used to select operators which describe the effective action of a scalar descending from a consistent theory of quantum gravity among all possible supersymmetric operators. In Section \ref{sec:global} we consider the structure of globally supersymmetric operators and briefly discuss the coupling to supergravity in Section \ref{sec:modstab}. 

%

\subsection{Higher-derivative operators in global supersymmetry}
\label{sec:global}

As outlined in the Introduction, our aim is to write \eqref{dbiresult} in the supersymmetric form
\begin{align}\label{action1}
\mathcal L = \int d \theta^2  d \bar \theta^2 K(\Phi,\bar \Phi) + \left( \int  d \theta^2 W( \Phi) + \text{h.c.} \right)\,,
\end{align}
where $\Phi = \phi + i \theta \sigma^\mu \bar \theta \partial_\mu \phi + \theta^2 F + \frac12 \theta^2 \bar \theta^2 \Box \phi$ denotes a chiral multiplet with its fermionic component set to zero.\footnote{We adopt the superspace conventions of \cite{Wess:1992cp}.} Hence we must find suitable higher-derivative terms to include in $K$.\footnote{The connection between the DBI action and higher-derivative supersymmetry or supergravity was previously studied in \cite{Khoury:2010gb,Koehn:2012ar,Koehn:2012np}. However, the previous analyses considered the kinetic terms for only one of the real scalars of the complex position modulus, freezing the other. This simplifies the discussion but leads to different results compared to the general case considered here. For a different approach, cf.~\cite{Tseytlin:1999dj,Sasaki:2012ka,Abe:2015nxa,Abe:2015fha}.} A list of operators with the desired amount of fields and derivatives was proposed in \cite{Khoury:2010gb}. A specific linear combination of these was singled out in \cite{Khoury:2010gb,Ciupke:2015msa},
\begin{align}\label{eq:op1}
\frac{1}{16} D \Phi D \Phi \bar D \bar \Phi \bar D \bar \Phi |_{\theta^2 \overline \theta^2} = |\partial_\mu \phi \partial^\mu \phi|^2  - 2 |F|^2 \partial_\mu \phi \partial^\mu \bar \phi + |F|^4\,.
\end{align}
The derivatives on the left-hand side denote the usual spinor-covariant derivatives. This term was deemed ``clean" in the sense that it is ghost-free and it contains no derivatives for the auxiliary field $F$. In addition, once the spinor is set to zero the operator is a pure D-term with no lower-order superspace components, and it contains only a single four-derivative term for $\phi$.

It is clear from the discussion in Section 2 that the structure in \eqref{eq:op1} is not what we find in the effective action of D-branes in string theory. While the first two pieces are indeed contained in \eqref{dbiresult} after identifying $V = |F|^2$, the term proportional to $|F|^4$ is not. More concretely, the term $|F|^4$ cannot be set to zero while, at the same time, keeping the correction to the kinetic term in the action. Therefore it cannot describe the particular cases studied in the previous section. This leads us to consider a number of other possible higher-derivative operators, focussing on those without a piece proportional to $|F|^4$, while postponing the discussion of unwanted states such as propagating auxiliary fields. To this end, the list of operators given in \cite{Khoury:2010gb} is particularly instructive. The relevant operators can be written in terms of component bosonic fields as follows, cf.~(8)-(13) in \cite{Khoury:2010gb},
\begin{align}
\mathcal O_1=|\Phi|^2 D^2 \Phi \bar{D}^2 \bar \Phi =\;\;& 16 |\phi |^2\Box \phi  \Box \bar \phi + 20 |F|^2\bar \phi \Box \phi + 20 |F|^2\phi  \Box \bar \phi  +16 |F|^4-8|F|^2\partial_{\mu}\phi  \partial^{\mu}\bar \phi  \nonumber \\
& +4|\phi |^2 F \Box \bar F  + 4 |\phi |^2 \bar F  \Box F - 8 |\phi |^2 \partial_{\mu}F\partial^{\mu}\bar F +8\bar \phi  F \partial_{\mu}\phi  \partial^{\mu}\bar F  \nonumber \\
&-8\bar \phi  \bar F  \partial_{\mu}\phi  \partial^{\mu}F+8\phi \bar F  \partial_{\mu}   \bar \phi  \partial^{\mu}F -8\phi F \partial_{\mu}\bar \phi  \partial^{\mu}\bar F \,, \label{listop1} 
\\
\mathcal O_2=\ \bar \Phi  \bar{D}^2 \bar \Phi (D\Phi )^2 =\;\;& 16 \partial_\mu \phi \partial^\mu \phi \bar \phi  \Box \bar \phi - 16|F|^2\bar \phi  \Box \phi + 16 |F|^2 \partial_\mu \bar \phi  \partial^\mu \bar \phi - 16 |F|^4 \nonumber \\
& + 16  \bar \phi  \bar F  \partial_\mu \phi \partial^\mu F  - 16 \bar \phi  F \partial_\mu \phi \partial^\mu \bar F   \,, \label{listop2}
\\
\mathcal O_3=|\Phi|^2 D \bar D \bar \Phi \bar D D \Phi =& \ 8(\partial_\mu \phi  \partial^\mu \bar \phi )^2 + 8\phi  \partial_\mu \bar \phi  (\partial_\nu \bar \phi  \partial^\mu \partial^\nu \phi  - 8\partial_\nu \phi  \partial^\mu \partial^\nu \bar \phi ) \nonumber \\
&- 8 |\phi |^2 \partial_\mu \phi  \partial^\mu \Box \bar \phi - 8|\phi |^2 \partial_\mu F \partial^\mu \bar F  - 8|F|^2 \partial_\mu \phi  \partial^\mu \bar \phi \nonumber \\
&- 8\bar \phi  F \partial_\mu \phi  \partial^\mu \bar F  - 8\phi  \bar F  \partial_\mu \bar \phi  \partial^\mu F\,, \label{listop4}
\end{align}
\begin{align}\mathcal O_4=\Phi^2 D \bar{D} \bar \Phi D \bar {D} \bar \Phi  =\;\;& -4 |\partial_\mu \phi  \partial^\mu \phi |^2 -4 \phi  \Box \phi  \partial_\mu \bar \phi  \partial^\mu \bar \phi -4 \phi ^2 \partial_{\mu}\partial_{\nu}\bar \phi  \partial^{\mu}\partial^{\nu}\bar \phi  \nonumber \\
 & -16 \phi  \partial_\mu \phi  \partial_\nu \bar \phi  \partial^\mu \partial^\nu \bar \phi - 4 \phi ^2 \partial_\mu \Box \bar \phi  \partial^\mu \bar \phi  - 32 \phi  F \partial_\mu \bar \phi  \partial^\mu \bar F\ .
 \end{align}
These are dimension-eight operators which in the action appear divided by $\Lambda^4$, where $\Lambda$ is the cut-off scale of the theory. In addition there are the complex conjugates $\overline{\mathcal O}_2$ and $\overline{\mathcal O}_4$. Notice that we did not include (14) and (15) of \cite{Khoury:2010gb} because, after partial integration, they are proportional to $\mathcal O_4$ and $\overline{\mathcal O}_4$, respectively. Important for us is that $\mathcal O_3$, $\mathcal O_4$, and $\overline{\mathcal O}_4$ span a basis of $|F|^4$-free operators. In particular, any $|F|^4$-free linear combination of $\mathcal O_1$, $\mathcal O_2$, and its complex conjugate can be expressed in terms of this basis. The operator \eqref{eq:op1}, on the other hand, is not described by this basis but is instead given by the linear combination
\begin{align}
D \Phi D \Phi \bar D \bar \Phi \bar D \bar \Phi = 2 \mathcal O_3 - \mathcal O_1 - \mathcal O_2 - \overline{\mathcal O}_2 \,.
\end{align}
While comparing supersymmetric operators to the DBI action one has to keep in mind that the latter does not capture higher-derivative contributions involving multiple derivatives of the scalar fields, for example terms containing $\Box \phi$ and $\partial_\mu \partial_\nu \phi$. Ignoring these we obtain
\begin{align}
\frac{\mathcal O_3}{\Lambda^4} &=  \frac {8}{\Lambda^4} \left[ (\partial_\mu \phi  \partial^\mu \bar \phi )^2  
- |\phi |^2 \partial_\mu F \partial^\mu \bar F  - |F|^2 \partial_\mu \phi  \partial^\mu \bar \phi  - \bar \phi  F \partial_\mu \phi  \partial^\mu \bar F  - \phi  \bar F  \partial_\mu \bar \phi  \partial^\mu F\right]\,, \label{listop4} \\
\frac{\mathcal O_4}{\Lambda^4} &= - \frac {4}{\Lambda^4} \left[  |\partial_\mu \phi  \partial^\mu \phi |^2  - 8 \phi  F \partial_\mu \bar \phi  \partial^\mu \bar F\right]\, .
\end{align}
Partial integration of the quartic kinetic terms introduces an ambiguity here, since terms with second derivatives can be written as first derivatives and vice versa. This ambiguity is manifest in a free coefficient of the four-derivative terms in the two expressions above. This makes the quartic kinetic terms not meaningful in the comparison with the DBI action. Thus, the strongest constraint on possible operators is indeed the absence or presence of $|F|^4$. All operators without $|F|^4$ can be written as
\begin{align}\label{genopresult}
c_1 \mathcal O_3 + c_2 \left(\mathcal O_4 + \overline{\mathcal O}_4 \right)\,.
\end{align}

\begin{figure}[t]
\centering
\includegraphics[scale=0.25]{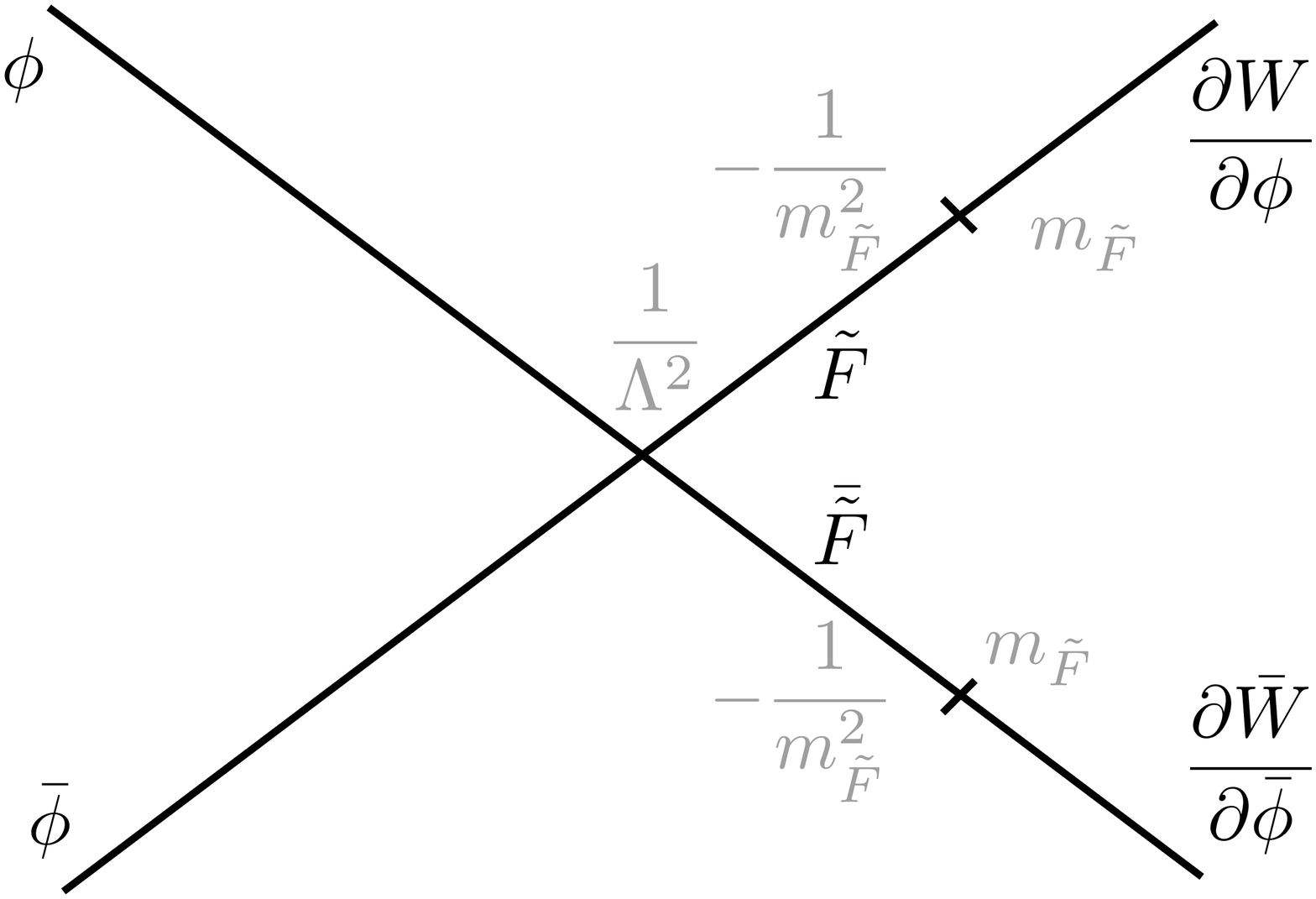}
\caption{Feynman diagram that leads to the presence of \eqref{vertex1} in the effective action.}
\label{fig:feynman1}
\end{figure}

Therefore this includes all operators that, after partial integration, yield \eqref{dbiresult} up to terms containing derivatives of $F$. Such terms seem to imply that the auxiliary field propagates. This would be unacceptable since we know from the DBI side that no such extra bosonic fields should be present. In fact, as emphasized in \cite{Baumann:2011nm}, derivative terms of auxiliary field are artefacts of the effective field theory description. Theories with higher-derivative corrections like \eqref{genopresult} must be UV completed above the cut-off scale $\Lambda$. The momenta of auxiliary fields with kinetic terms from higher-derivative operators are larger than $\Lambda$ and are hence irrelevant in the EFT. This argument is strongly supported by the fact that  UV-complete theories, such as the DBI action, should be  free of ghosts and propagating auxiliary fields. To see this more explicitly, note that the lowest-dimensional action in \eqref{action1} contains the bosonic pieces
\begin{align}
\mathcal L \supset - |F|^2 - \left(F \frac {\partial W}{\partial \phi} +\text{h.c.} \right) \, .
\end{align}
To obtain the standard mass dimension for the field $F$ we redefine ${\tilde F}=F/\Lambda$. We thus get
\begin{align}
\mathcal L \supset  -m^2_{\tilde F} |{\tilde F}|^2 - m_{\tilde F} \left( {\tilde F} \frac {\partial W}{\partial \phi}+ \text{h.c.} \right)\,,
\label{truco}
\end{align}
with $m_{\tilde F} = \Lambda$. Thus, actually the scalar field ${\tilde F}$ has a mass of the same order as the cut-off scale and should decouple below the scale $\Lambda$. One has to be careful though, since  integrating out ${\tilde F}$ is not equivalent to setting $m_{\tilde F} \to \infty$, due to the presence of the dimensionful coupling of ${\tilde F}$ to $\phi$ in the above expression. In an effective action description one neglects all terms proportional to $\partial_\mu {\tilde F} /m_{\tilde F}^2$. This leads us to conclude that, ignoring the quartic kinetic terms, the operators $\mathcal O_4$ and $\overline{\mathcal O}_4$ above may be ignored and the operator $ \mathcal O_3$ is left with the only desired piece
\begin{align} 
\mathcal O_3 = - \frac {8}{\Lambda^2} |{\tilde F}|^2 \partial_\mu \phi  \partial^\mu \bar \phi  + \mathcal O \left((\partial_\mu \phi)^4\right)  \, .
\label{ello}
\end{align}
One might be tempted to argue that not even this term survives in the effective action because ${\tilde F}$ decouples. However, it is easy to convince oneself that this is not the case due to the second term in \eqref{truco}. Indeed, as shown in Figure~\ref{fig:feynman1}, one can draw a tree-level Feynman diagram with a vertex stemming from \eqref{ello} and two ${\tilde F}$ propagators. The latter end in vertices provided by the second piece in \eqref{truco}. In the effective action limit with $(\partial_\mu {\tilde F})\ll m_{\tilde F}^2$ the propagator of ${\tilde F}$ is approximately $-1/m_{\tilde F}^2$ so that, in the end, we are left with
\begin{align}\label{vertex1}
\mathcal O_3 = - \frac {8}{\Lambda^4} \left| \frac {\partial W}{\partial \phi}\right|^2 \partial_\mu \phi  \partial^\mu \bar \phi  + \mathcal O \left((\partial_\mu \phi)^4\right) \, .
\end{align}
In conclusion we find the effective action
\begin{align}
\mathcal L = - \left(1 + \frac {8c_1}{\Lambda ^4} \left| \frac {\partial W}{\partial \phi}\right|^2 \right)  \partial_\mu \phi \partial^\mu \bar \phi - \left| \frac {\partial W}{\partial \phi}\right|^2 + \mathcal O \left(\Box \phi, \partial_\mu \partial_\nu \phi, (\partial_\mu \phi)^4\right) \,,
\label{effectiveaction}
\end{align} 
which derives from the supersymmetric action
\begin{align}
\mathcal L = \int d\theta^2d {\bar \theta}^2  |\Phi|^2 \left(1 + \frac{c_1}{\Lambda^4}D \bar D \bar \Phi \bar D D \Phi \right) + \left( \int d\theta^2 W(\Phi) +\text{h.c.} \right) \,.
\end{align}
Using the identity $D_\alpha \bar D_{\dot \alpha} \bar \Phi = \{D_\alpha, \bar D_{\dot \alpha}\} \bar \Phi = -2i\sigma^\mu_{\alpha \dot \alpha} \partial_\mu \bar \Phi$ we can write this action in the more transparent fashion
\begin{align}\label{hdkahler}
\mathcal L = \int d\theta^2d {\bar \theta}^2 |\Phi|^2 \left(1+ \frac{8c_1}{\Lambda^4} \partial^\mu \Phi \partial_\mu \bar \Phi\right) + \left(\int d\theta^2 W(\Phi) +\text{h.c.} \right)  \, .
\end{align}
Note that, once coupled to $\mathcal N=1$ supergravity, the auxiliary field ${\tilde F}$ should be replaced by the corresponding chiral multiplet auxiliary field of supergravity. In \eqref{effectiveaction} this implies that one should replace field derivatives by K\"ahler-covariant derivatives, $\partial_\phi W \to D_\phi W$. Before describe the coupling to supergravity in more detail, one more comment regarding D-terms is in order.

While studying the D$p$-brane DBI actions in Section 2 we noticed that the D-term potential multiplies kinetic terms as well. We set all D-terms to zero so that supersymmetry is preserved and concentrated on the analysis of the more interesting F-term scalar potential. However, it is interesting to find the corresponding higher-derivative operator including D-terms as well. In the simple example above there is a single adjoint multiplet with the standard gauge transformations
\beq 
\Phi \to e^{-i\Lambda (x)} \Phi\,, \qquad e^V \to e^{-i{\bar \Lambda}(x)} e^V e^{i\Lambda (x)} \,,
\eeq
where $V=T^aV_a$ denotes the vector multiplet and $\Lambda=T^a\Lambda_a$. $T^a$ are the gauge group generators and $\Lambda_a(x)$ the gauge parameter superfields. Then the operator $\mathcal O_3$ can be generalized to the gauge-invariant operator
\beq
{\tilde { \mathcal O_3}} = ({\bar \Phi} e^V \Phi)( D \bar D  {\bar \Phi} e^V)(\bar D D e^V \Phi ) \, .
\eeq
Expanding the $\theta^2 {\bar \theta}^2$ component of this operator one obtains a coupling of the form
\beq
\frac {8}{\Lambda^4} (\bar \phi D_aT^a\phi){\cal D}_\mu \phi {\cal {\bar D}}^\mu \bar \phi \, ,
\eeq
where $D$ is the auxiliary field $D=T^aD_a$, and $\cal D$ is the standard gauge-covariant derivative. Using the equations of motion for $D$ one obtains $D_a=-\bar \phi T^a\phi$. In the above expression this yields the familiar structure
\beq
-\frac {8}{\Lambda^4}  V_D {\cal D}_\mu \phi {\cal {\bar D}}^\mu \bar\phi  \, ,
\eeq
where $V_D$ is the standard D-term scalar potential. These formulae apply, for example, to the toroidal D7 case discussed above, which has a single adjoint position modulus. The  generalization to the case of multiple chiral superfields is once more straightforward. 

Note that in the D7-brane example there are still more adjoint scalars from the Wilson lines. These would give rise to additional pieces in the D-term potential which do not depend on the position moduli. Finally, let us note that the DBI action contains terms which rescale the inverse
gauge coupling constant, i.e., terms proportional to $V(\phi)F_{\mu \nu}F^{\mu \nu }$, as the reader can easily check. Those can be described by the supersymmetric operators
\beq
\frac {1}{\Lambda ^4} ({\bar \Phi} e^V\Phi) D^2W^\alpha W_\alpha +\text{h.c.} \,,
\eeq
where $W_\alpha$ is the spinorial gauge field-strength which admits an expansion $W_\alpha = \lambda_\alpha +(\sigma ^\mu {\bar \sigma}^\nu \theta)_\alpha  F_{\mu \nu }+ \dots$.
The D-term component of such an operator defines the non-trivial kinetic term of the gauge bosons. This behavior is expected because if the potential increases, so does the tension of the brane, which in turn implies a smaller gauge coupling. Thus, for example in large-field inflation models from D-branes the corresponding gauge coupling decreases with increasing inflaton field value.

The operators found in this section are interesting since they tells us how to embed the non-canonical kinetic terms of the string theory DBI action into a supersymmetry or supergravity action. As emphasized before, it can be easily generalized to the case of multiple scalar fields which may appear in D$p$-brane configurations in different compactifications. Moreover, an important conclusion is that the higher-derivative corrections implied by the DBI action do not include terms proportional to $|F|^4$, which can lead to multiple vacuum configurations. They instead imply a non-canonical redefinition of the kinetic terms proportional to the scalar potential and higher-derivative kinetic terms. In the next section we discuss how these corrections appear in a supergravity setting.

%

\subsection{$\mathcal N =1$ supergravity description}
\label{sec:modstab}

The generalization of the previous findings to local supersymmetry can be done along the lines of \cite{Baumann:2011nm}. Indeed, it can be shown that the K\"ahler potential in \eqref{hdkahler} produces the same effective scalar field theory when coupled to gravity, after the supergravity auxiliary fields have been put on-shell. This has a number of important implications for the study of inflationary models involving D-brane position moduli. 

An $\mathcal N=1$ supergravity description of the effective theory for a D-brane position modulus is desirable in order to study the consequences of closed string moduli stabilization. The interaction between the dynamical closed string modes and the open string inflationary sector is not captured by the DBI and CS actions. However, as of now such a supergravity formulation would miss the flattening effect of the non-canonical kinetic term of $\Phi$, due to $\alpha'$ corrections in the DBI action which are not visible in standard two-derivative supergravity. With the results of Section \ref{sec:global}, in particular \eqref{hdkahler}, we can now capture this effect in supergravity. 

For concreteness, let us focus again on a single chiral superfield corresponding to the position modulus of a D7-brane in a toroidal setting, as in the Higgs-otic example discussed in
\cite{Ibanez:2014swa}. The relevant piece of the K\"ahler potential in this class of
Type IIB orientifold compactifications with D7-branes is, at leading order in $\alpha'$, given by \cite{LopesCardoso:1994is,Antoniadis:1994hg,Brignole:1995fb,Brignole:1996xb,Ibanez:2014swa}
\begin{align}
K &= -\log\left[(S+\bar S) (U + \bar U) - \frac12 (\Phi + \bar \Phi)^2\right] -3 \log[T + \bar T]\, , \\
W &= W_0 + \mu \Phi^2 \,,\label{KW}
\end{align}
where $S$, $T$, and $U$ denote the axio-dilaton, an overall K\"ahler modulus, and a complex structure modulus of the third torus, respectively. Notice the shift-symmetric structure of the K\"ahler potential for the position modulus contained in $\Phi$, which leads to an approximate continuous shift symmetry in the scalar potential broken by fluxes. This flat direction in the K\"ahler potential is not only present in toroidal compactifications, but also in generic Calabi-Yau compactifications in the large complex structure limit, and it is expected to be preserved by all perturbative corrections to $K$. Assuming that the potential is minimized when $D_S W = D_U W = 0$, the dominant source of supersymmetry breaking is the auxiliary field of $T$, which leads to a soft mass for the D7 matter field $\Phi$. Both contributions $W_0$ and $\mu$ in the superpotential are required to match the DBI result \eqref{VD7} with non-vanishing $G$ and $S$ fluxes. The precise matching at leading order in $\alpha'$ was worked out in \cite{Ibanez:2014swa}. The novel feature here is the addition of the higher-derivative piece, which can be written as a correction of the K\"ahler potential given by
\beq
\Delta K=\frac{1}{(S+\bar S) (U + \bar U) }\frac{8c_1}{\Lambda^4} \left[ (\Phi+\bar\Phi)^2 \partial_\mu \Phi \partial^\mu \bar \Phi  \right]\,.
\label{deltaK}
\eeq
We have used a variation of the result in \eqref{hdkahler} to keep the shift symmetry manifest in the K\"ahler potential. The same higher-derivative operator was previously studied in \cite{Baumann:2011nm}. After integrating out the auxiliary field and ignoring the quartic kinetic terms the result is equivalent to \eqref{effectiveaction}. The scaling with the axio-dilaton and the complex structure moduli is required by modular invariance of $K$. This extra piece leads to the correction of the kinetic term as in \eqref{higgsotic2} after identifying $a=8c_1/\Lambda^4$. Using the result for $a$ derived from the DBI D7-brane action, we find
\beq
\frac{c_1}{\Lambda^4}=(8V_4 \mu_7 g_s)^{-1}\simeq \frac{\pi\alpha_G}{2 g_s M_s^4}\,,
\eeq
where $\alpha_G$ is the gauge coupling and $M_s=\sigma^{-1/2}$ the string scale. This makes the stringy nature of the higher-derivative correction manifest. In terms of the Planck mass we obtain $c_1/\Lambda^4 = 16\pi^3 / ( \alpha_G^2 M_\text{p}^4)$. For the last two expressions we have used the typical IIB identities \cite{Ibanez:2012zz}
\beq
8\pi M_\text{p}^2=\frac{8M_s^8V_6}{(2\pi)^6g_s}\label{Mp}\,,\qquad 
M_\text{KK}=M_s\left(\frac{2\alpha_G}{g_s}\right)^{1/4} \, ,
\eeq
where $V_6$ denotes the volume of the internal space and $M_\text{KK}=V_6^{-1/6}$ the compactification scale. 

As a consequence, the above analysis is a step towards a complete supergravity formulation of the DBI action. It permits us to study the interplay between open string modulus dynamics and moduli stabilization while taking into account the flattening of the quadratic flux potential by the non-canonical kinetic term. This is of particular interest for inflation models in which the inflaton is a D-brane position modulus like in Higgs-otic inflation \cite{Ibanez:2014swa}. While this is very appealing, great care is needed when analyzing such setups. On the one hand, the interaction between closed-open string moduli and the resulting coupling terms in the superpotential are model-dependent and not completely known in general. On the other hand, as discussed in \cite{Buchmuller:2015oma,Dudas:2015lga}, the interaction of supersymmetry-breaking closed string moduli with inflation is non-trivial and can cause numerous types of trouble. Hence we leave the details of the study of moduli stabilization for future work. 


\section{Flattening of inflationary potentials}
\label{sec:inflation}

In this section we analyze the effect of the DBI non-canonical kinetic term on inflationary observables. We assume that one of the real components of $\phi$ is the inflaton field which has a potential suitable for slow-roll. Moreover, we work in the slow-roll regime and thus neglect the fourth-order derivative term of $\phi$ in \eqref{dbiresult}. What we study is therefore a version of the Lagrangian \eqref{higgsotic2} with a single real scalar field $\varphi$,
\begin{align}
\mathcal{L}  = -\frac{1}{2} f(\varphi) \partial_\mu \varphi \partial^\mu \varphi - V(\varphi)\,,
\end{align}
where
\begin{align}
f(\varphi) = 1+aV(\varphi)\,.
\label{eq:startingDefs}
\end{align}
The effect of taking both degrees of freedom of the complex field $\phi$ into account was studied in \cite{Bielleman:2015lka} for the case of D7-branes. In Section \ref{sec:DBI} we found that the parameter $a$ for the case of a D$p$-brane is $a \sim (\mu_p V_{p-3})^{-1}$. This implies that $a$ is of the order $M_s^{-4}$, encoding the stringy nature of the correction.

As emphasized above, the DBI action yields a non-canonical kinetic term for the inflaton. However, in single-field inflation models one can always recast the Lagrangian into a canonical form via a field redefinition. The proper redefinition is determined by the differential equation
\begin{align}
\frac{d \varphi}{d \psi}=\frac{1}{f^{1/2}(\varphi)}=\frac{1}{\sqrt{1+a V(\varphi)}} \,,
\end{align}
which yields
\begin{align}
\psi =  g(\varphi) = \int d\varphi f^{1/2}(\varphi) \,.
\label{eq: fieldredef}
\end{align}
The Lagrangian, when written in terms of the canonically normalized field $\psi$, reads 
\begin{align}
\mathcal{L}  =  - \frac{1}{2}\partial_\mu \psi \partial^\mu \psi -V(g^{-1}(\psi)) \,,
\end{align}
so that $V$ implicitly depends on $a$. Interestingly, this process leads to a flatter potential. Specifically,
\begin{align}
\frac{\partial V}{\partial \psi}  = \frac{1}{f^{1/2}} \frac{\partial V}{\partial \varphi} \, .
\end{align}
Since $f >1$ if $a>0$ the potential in canonical variables has a smaller first derivative, i.e., a flattened slope. A similar flattening from non-canonical kinetic terms has been discussed in the past in the context of string cosmology, for example in \cite{Gur-Ari:2013sba}.

\begin{figure}[t]
\centering
\includegraphics[scale=0.5]{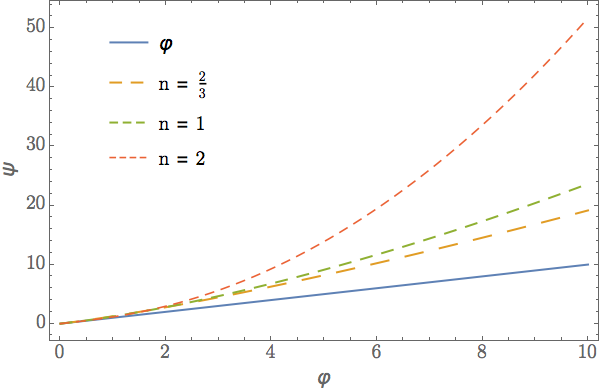}
\caption{$\psi(\varphi)$ for monomial potentials of various powers $n$.}
\label{fig:chaotic}
\end{figure}

Provided $f>0$, i.e., the scalar field is not a ghost, the study of the vacua can be performed by analyzing $V(\varphi)$ and neglecting the non-canonical nature of the field. The dynamics of the theory, however, crucially depend on the redefinition of the kinetic term. To quantify this effect we compute the CMB observables in terms of the canonically normalised field, first as general as possible and later applied to monomial potentials. We define the potential slow-roll parameters as usual,
\begin{align}
\epsilon = \frac{1}{2} \left(\frac{V_\psi}{V} \right)^2\,, \qquad \eta=\frac{V_{\psi\psi}}{V}\,,
\end{align}
where subscripts denote derivatives. These can be rewritten in terms of $\varphi$ as follows,
\begin{align}
\epsilon = \frac{1}{2f} \left(\frac{V_\varphi}{V} \right)^2\,, \qquad \eta=\frac{1}{f} \frac{V_{\varphi\varphi}}{V}- \frac{a  V}{f}\epsilon \, .
\end{align}
Evidently, the effect of the non-canonical kinetic terms is to reduce the slow-roll parameters. The scalar spectral index of the curvature perturbations is
\begin{align}
n_\text{s}&=1-6 \epsilon+2\eta \, , \nonumber \\
&=1-\frac{3}{f}\left(\frac{V_\varphi}{V}\right)^2+\frac{2}{f}\frac{V_{\varphi\varphi}}{V}-\frac{aV}{f^2}\left(\frac{V_\varphi}{V}\right)^2\, \nonumber, \\
&= \frac{1}{f} (1-6 \epsilon |_{a=0}+2 \eta |_{a=0}) + \frac{aV}{f}(1-2 \epsilon) \, ,
\label{eq:ns}
\end{align}
where in the last line only the second piece depends on $a$. The tensor-to-scalar ratio becomes 
\begin{align}
r=16\ \epsilon=\frac{8}{f} \left(\frac{V_\varphi}{V}\right)^2.
\label{eq:r}
\end{align}
Both $n_\text{s}$ and $r$ are to be evaluated at horizon exit, with field values denoted by $\psi_*$ and $\varphi_*$. For $N_e$ $e$-folds of exponential expansion one has
\begin{align}
N_e=\int_{\psi_\text{end}}^{\psi_*}\frac{1}{\sqrt{2 \epsilon} }d\psi= \int_{\varphi_\text{end}}^{\varphi_*}f\frac{V}{V_\varphi} d \varphi \,,
\label{eq: efolds}
\end{align}
which defines $\varphi_*$ and $\psi_*$. The difference between $\varphi_*$ and $\psi_*$ and $\varphi_\text{end}$ and $\psi_\text{end}$, respectively, is model-dependent. Therefore, in the following, we study simple examples and quantify the effect of the non-canonical normalization numerically. As discussed in Section \ref{sec:DBI}, world-volume and background fluxes generate monomial potentials for D-brane position moduli. We therefore consider potentials of the type
\begin{align}
V_n(\varphi)= v_0\ \varphi^n\,,
\end{align}
with $n \in \mathbb{R}_+$. In this case we can specify $g(\varphi)$ in \eqref{eq: fieldredef},
\begin{align}
\psi = \frac{\varphi\left[2\sqrt{1+a v_0 \varphi^n}+n\ _2F_1(\frac{1}{2},\frac{1}{n};1+\frac{1}{n};-a v_0\varphi^n)\right]}{2+n}\, ,
\label{eq:chaoticFieldRedef}
\end{align}
where $_2F_1(a,b;c;d)$ is the ordinary hypergeometric function. Note that $\psi$ is real only when $-a v_0\varphi^n < 1$ which is equivalent to the no-ghost regime. We illustrate the functional dependence of \eqref{eq:chaoticFieldRedef} in Figure~\ref{fig:chaotic} for representative values of $n$.
The crucial feature of this plot is that all curves lie above the $\psi=\varphi$ reference line, implying that one may schematically write $\psi = \varphi^{m(\varphi)}$ for some $m(\varphi)>1$, or equivalently $\varphi=\psi^{1/m(\psi)}$. Since $m(\psi)>1$ the change to a canonically normalized inflaton results in a monomial potential with suppressed power. The schematic form is $V\sim \psi^{n/m(\psi)}$, demonstrating that the effect of the non-canonical coupling in \eqref{eq:startingDefs} is to cause a flattening of the monomial potential. Given the monotonicity of the scalar potentials we thus expect $n_\text{s}$ to increase while $r$ decreases.

While a proper field redefinition exists for all $n$ there are only a few values for which we can use functional identities to rewrite \eqref{eq:chaoticFieldRedef} in a more familiar form,
\begin{align}
n=0 &: \quad \psi=\sqrt{1+av_0}\varphi + C \, , \\
n=1 &: \quad \psi = \frac{2}{3av_0} \left[(1+av_0\varphi)^{3/2}-1 \right] +C \, , \\
n=2 &: \quad \psi=\frac12 \varphi \left[\sqrt{1+av_0\varphi^2}+\frac{1}{\sqrt{av_0}\varphi} \text{ arsinh}(\sqrt{av_0}\varphi) \right]+C\,,
\end{align}
with $C=0$ fixed by the requirement $V(0)=0$, i.e., demanding the cosmological constant to vanish in the vacuum. Notice that, as expected, in the first case of a trivial potential the field redefinition is simply a rescaling of $\varphi$. But because $V$ is constant this is also the most uninteresting case. On the other hand, for $n=2$ as in single-field Higgs-otic inflation there is no analytical form for the inverse $\varphi(\psi)$. This exists only for $n=1$. This means that, to study the implications in the most interesting cases, we must resort to either approximations or numerics. In the remainder of this section we use a combination of both.

\begin{figure}[t]
\centering
\includegraphics[scale=0.35]{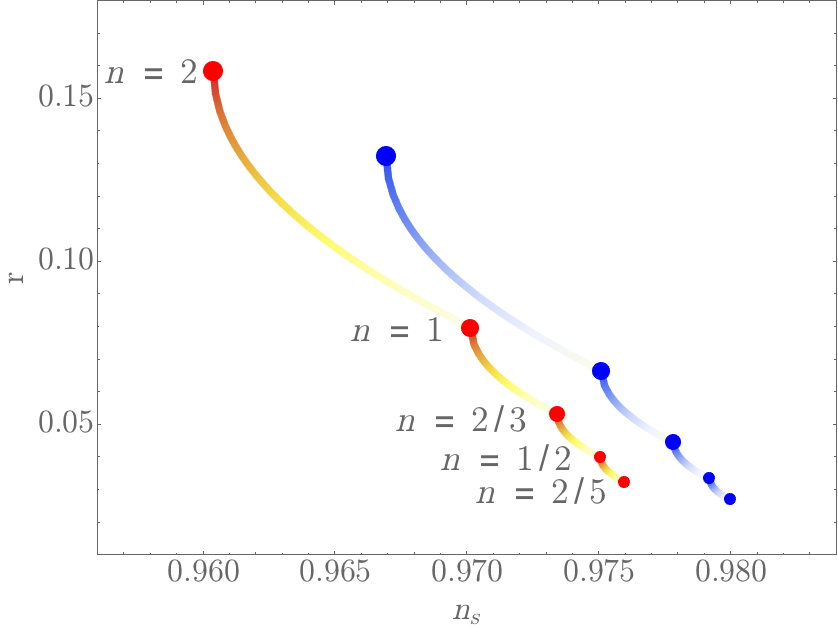}
\caption{CMB observables as predicted by the canonically normalized theory, with initial values $n=2$, $n=1$, $n= \frac23$, and $n=\frac23$. Darker color means larger values of $av_0$. For small $a v_0$ the effect of the additional kinetic term is negligible, while for large $a v_0$ the potential $V(\psi)$ approaches a monomial with power $1$ for $n=2$, $\frac23$ for $n=1$, $\frac12$ for $n=\frac23$, and so on. The two distinct lines correspond to $N_e = 50$ and $N_e = 60$, respectively.}
\label{fig: chaotic}
\end{figure}

We present the results of a numerical analysis of the CMB observables in Figure \ref{fig: chaotic}, using $n=2,1,\frac23, \frac25$ as examples.
We vary the value of $av_0$ to study the strength of the flattening effect. Remember that increasing $a v_0$ means power suppression in the monomial potential of the canonically normalized inflaton. To better understand the numerical results let us first consider the limit of small $a v_0$, so that $a V \ll1$ and $f \simeq 1$ in \eqref{eq:startingDefs}. A first-order Taylor expansion in $a v_0$ leads to simplified expressions for the slow-roll parameters,
\begin{align}
\epsilon &= \frac12 (1- av_0 \varphi^n)\frac{n^2}{\varphi^2} \, ,\\
\eta &=  \left(1-av_0 \varphi^n \right) \left( \frac{n(n-1)}{\varphi^2}- \frac{n^2}{2}av_0\varphi^{n-2}\right) \, .
\end{align}
For $\varphi_\text{end}$, defined by $\epsilon(\varphi_\text{end})=1$, we find
\begin{align}
\varphi_\text{end}  = \frac{n}{\sqrt{2}}\left (1-\frac{a v_0 }{2}\left(\frac{n}{\sqrt{2}}\right)^n\right) \, .
\end{align}
Furthermore, the observable modes of the fluctuations leave the horizon at
 \begin{align}
\varphi_* =  \sqrt x -\frac{av_0}{n+2}\left(x^{\frac{n+1}{2}} + (n+1) \left(\frac{n}{\sqrt{2}}\right)^n \sqrt x\right) \, ,
\label{eq:phiStar}
\end{align}
where we have introduced $x = 2n N_e +\frac12 n^2$. Using this value in the expanded slow-roll parameters leads to
\begin{align}
\epsilon_*  &=  \frac{n^2}{2x} + av_0 n^2 \left(\frac{n+1}{n+2} \left(\frac{n}{\sqrt{2}}\right)^{n+2}x^{-2}-\frac{n}{2n+4} x^{\frac12 n-1} \right)\,, \\
\eta_*  &=  \frac{n^2-n}{x} +a v_0n \left(\frac{2n^2+2}{n+2} \left(\frac{n}{\sqrt{2}}\right)^{n+2} x^{-2} -\frac{3n^2}{2n+4} x^{\frac12 n-1}\right)\,,
\end{align}
at horizon exit. The structure is remarkably similar in both cases, which can be traced back to the term proportional to $aV/\epsilon$ in the integral that determines $N_e$. We expect the term proportional to $x^{\frac12 n-1}$ to dominate in the brackets because $x \sim \mathcal O(100)$. Hence, both functions decrease as $a v_0$ increases. In the limit of small $a v_0$ this explains why the observables move towards the bottom-right in the $n_\text{s}$-$r$ plane as the non-trivial kinetic term is amplified. 

The limit of large $a v_0$ is even more illuminating, cf.~the related analyses in \cite{Gur-Ari:2013sba,Aoki:2015eba}. Assuming $f \simeq aV$ leads to
\begin{align}
\varphi &= \left( \frac{n+2}{2\sqrt{av_0}}\right)^{\frac{2}{n+2}}\psi^{\frac{2}{n+2}} \, ,
\end{align}
as the inverse of \eqref{eq:chaoticFieldRedef}. The corresponding scalar potential becomes
\begin{align}
V(\psi)  =  v_0 \left( \frac{n+2}{2\sqrt{av_0}}\right)^{\frac{2n}{n+2}}\psi^{\frac{2n}{n+2}} \, .
\end{align}
Thus, we obtain an analytic result for the canonically normalized theory for any value of $n$. In particular, starting with a power of $n$ in $\varphi$ we obtain a power of $\frac{2n}{n+2} < n$ in the canonical field $\psi$. This explains another feature in Figure~\ref{fig: chaotic}: in the regime of large $a$, starting with $n=2$ yields a monomial potential of power $1$, $n=1$ yields power $\frac23$, $n=\frac23$ leads to $V \sim \psi^\frac12$, and so on. This is why the curves in the figure connect.

%

\section{Conclusions}

In this paper we have studied the appearance of higher-dimensional supersymmetric operators correcting the K\"ahler potential in string theory. More concretely, we have studied the K\"ahler potential of D-brane position moduli in Type II orientifold compactifications, which arises from the Dirac-Born-Infeld action for D-branes. We have concentrated on the effective action for the position moduli of Type IIB D$p$-branes with $p=3,5,7$ on toroidal settings. One of the important conclusions is that in all cases dimension-eight corrections arise which induce non-canonical kinetic terms of the form $\left[1+aV(\phi)\right]\partial_\mu \phi \partial^\mu \bar \phi$, where $V$ is the scalar potential. Upon canonical normalization this implies a flattening of the scalar potential for large canonical field values. In specific backgrounds the potential $V$ is just the leading-order scalar potential $V_0$. In particular, no higher powers of $V_0$ appear in the effective action. We have identified the supersymmetric dimension-eight operators describing these purely stringy corrections. They have the superfield form $|\Phi|^2\partial_\mu \Phi \partial^\mu \bar \Phi$. Although this contains derivatives of the auxiliary field, there are no new propagating degrees of freedom or ghosts in the effective action once states with masses of the order of the cut-off scale are properly integrated out. Moreover, the above operator does not include the term $|F|^4$ unlike the operator $D\Phi D\Phi{\bar D} \bar \Phi{\bar D}\bar \Phi$ which has been studied in the past. 

The above results are interesting in themselves but also have important implications for string inflation models. They allow for an $\mathcal N=1$ supergravity description of string inflation models in which the inflaton is an open string D$p$-brane modulus. For example, \cite{Ibanez:2014swa} proposed a supergravity description of Higgs-otic inflation where the inflaton is a linear combination of the MSSM Higgs fields, without non-minimal couplings of the Higgs fields to gravity. In such a model  the Higgs fields are D7-brane position moduli and the scalar potential is defined in terms of a DBI action and the flux background. A flattening of the scalar potential takes place of the type described in the present paper. The supergravity version of the model in \cite{Ibanez:2014swa} did not include these flattening effects. Thus, the results in this paper allows us to complete the supergravity embedding, which may allow the detailed study of closed string moduli stabilization. 
 
Given the pervasive presence of the non-canonical kinetic term in \eqref{higgsotic} we found it interesting to explore its impact in simple single-field inflation models, like chaotic models with monomial potentials as suggested by orientifold compactifications with fluxes. We have studied the effect of the aforementioned flattening on the slow-roll parameters and the scalar and tensor perturbations. As expected, we found a suppression of the tensor-to-scalar ratio in all cases, leading to an improved agreement with present Planck and BICEP constraints. Furthermore, we have presented simple analytic formulae explaining how this takes place. Hopefully, forthcoming cosmological data will shed light on the existence of large primordial tensor perturbations, which could also illuminate the role of higher-dimensional operators in inflation models.

%

\section*{Acknowledgments}

We thank D. Baumann, F. Marchesano, and A. Westphal for useful discussions. This work is partially supported by the grants  FPA2012-32828 from the MINECO, the ERC Advanced Grant SPLE under contract ERC-2012-ADG-20120216-320421 and the grant SEV-2012-0249 of the ``Centro de Excelencia Severo Ochoa" Programme.  I.V. is supported by a grant from the Max Planck Society. 

%

\end{document}